# Possibility to make a beam of τ –leptons and charmed particles by a channeling crystal

V.M. Biryukov

Institute for High Energy Physics

Protvino, Russia

**Abstract.**

We suggest a way to capture by a focusing crystal the short-lived particles (tau leptons, charmed baryons) produced in the decays downstream of a target hit by primary protons. The beam of captured short-lived particles can be extracted from the debris produced in the target and bent onto experimental setup over a distance of centimetres. The debris is rejected by the crystal, i.e. it remains nonchannelled. The technique can handle the particles with decay length $c\tau$ down to 1 micron.

The technique of beam steering at particle accelerators by means of bent crystal channeling has matured from the idea of E.N. Tsyganov [1] to routine applications for beam delivery on everyday basis [2]. The technique started at a few GeV [3] and has expanded down to 3 MeV [4] and up into TeV range [5,6]. Starting from protons, it progressed to heavy ions, pions, positrons, and even short-lived particles [7-9]. More recently it was demonstrated also for negative particles (electrons, pions) [10]. Beams can be deflected and focused by channeling crystals [11]. Crystals can even form a beam line composed of a string of several crystal elements [12] just as the traditional magnet beam line made of dipoles and quadrupoles. The field strength in bent crystals is equivalent to a few hundred Tesla in routinely used Si crystals and can be up to a few thousand Tesla in heavier crystals such as Germanium [13]. Beams of about $10^{12}$ proton/s intensity were channeled by Si crystals with efficiency of 85% [14].

Several applications of the technique are discussed for the LHC: crystal collimation [15], beam extraction for fixed target physics [16], magnetic and electric moment measurements for short-lived particles (with charm and beauty quarks, and τ leptons) [17-20] to name just a few.

Below we consider the capabilities of crystal channeling technique to handle short-lived particles (mesons and baryons with charm and beauty quarks, and τ leptons) at the high energy frontier, LHC and FCC. The traditional magnet optics handled first the stable particles (protons, ions, electrons). Later, with the energy rise in accelerators, the magnet optics could form and transport the beams of short-lived strange particles (kaons) providing the opportunities for interesting experiments on kaon physics leading to fundamental discoveries. The energy rise gives a Lorentz boost to the mean path $\gamma c\tau$ of a short-lived particle in the laboratory frame. Here $\gamma$ is the Lorentz factor, $c$ the light speed, $\tau$ the lifetime of the particle. The lifetimes of strange particles are in the range of $10^{-10}$ to $10^{-7}$ s with $c\tau$ up to a few meters.

Short-lived particles such as charm and beauty mesons and baryons have the lifetime $\tau$ so short that $c\tau$ equals only 30-500 μm. This is 4-5 orders of magnitude less than $c\tau = 3.7$ m for kaons $K^{\pm}$, strange particles. Now at high energy frontier, instead of typical 1 GeV kaons the physicists have to handle 1 TeV baryons and leptons. Even with relativistic factor of 1000 the mean path before decay for these particles equals only 3-50 cm. There is no way to handle these particles by traditional magnet beam optics.

The three orders of magnitude increase in energy from GeV to TeV, and corresponding Lorentz boost in mean free path $\gamma c\tau$ of particles, is not enough to compensate for the 4-5 orders shortage in particles lifetime compared to kaons. However, the technique of beam handling by means of channeling crystals developed in the past decades offers us the fields for beam bending and focusing as strong as a few hundred (up to a few thousand) of Tesla. This is 2-3 orders of magnitude higher than the fields of traditional magnetic optics for beam steering.

We notice that together these two factors, three orders in the energy rise and two-three orders in the field strength, give us 5-6 orders which are enough to compensate for the 4-5 orders shortage in the lifetime of charm and beauty particles and tau leptons compared to kaons $K^{\pm}$. Whatever the accelerator scientists did with the strange particles, making beams of kaons, now should be possible to do with the short-lived particles of charm and beauty, and tau leptons.

Below we propose one idea how a channelling crystal could be used in order to make a beam of short-lived particles selected and trapped from the debris produced by a primary proton beam in a target and then bent (extracted from the production cone) onto an experimental setup.

The discussed idea involves a focusing crystal, which can either focus a parallel beam into a point, or perform a reversed process, i.e. trap a beam emerging from a point-like source; both applications have been demonstrated experimentally [11,12]. Such a crystal has a focal point where the parallel beam is focused to by the crystal, or – in a reversed process – the particles emerging from this focal point are trapped by the channeling crystal.

The method also involves a target bombarded by primary proton beam that produces all kinds of particles in the target. Some of the produced secondary particles are short-lived ones which decay shortly downstream of the target.

Suppose a target (to produce short-lived particles) and a focusing crystal are arranged so that the crystal focal point $F$ is outside of the target (Fig. 1). That is, a focusing crystal looks at the spot downstream of the target, but not at the target itself. As a result, the crystal traps the particles emerging from this spot but does not trap the particles emerging from the target. This way we select the decay products of the short lived particles, such as tau leptons, but we ignore the debris coming from the target. As only the particles emerging from $F$ can be trapped into channeling mode by the crystal, then, if some particle decays at $F$, the decay products will be trapped with some efficiency.

If some particle emerges from the target directly, it is not trapped, unless it passes through $F$. This consideration sets a "blind spot" $B$ on the crystal face, where one cannot distinguish between the particles of interest and the background. The trapped (channeled) particles can be bent, e.g. onto some experimental setup.

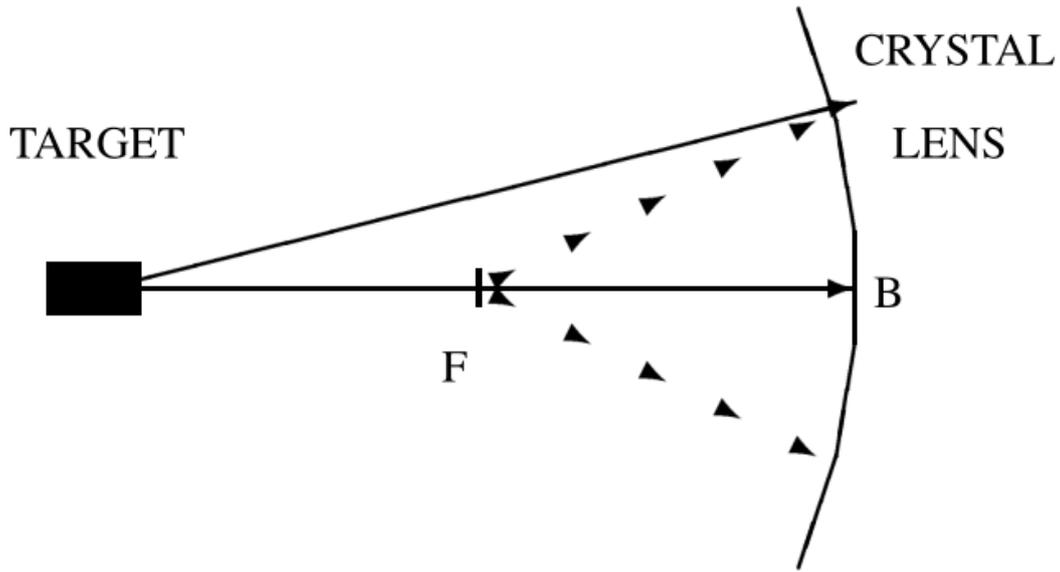

**Fig. 1** Schematic arrangement of a target, the entry face of a focusing crystal ("crystal lens"), and the focus F. "Blind spot" is also shown (B).

So, we need a primary (multi-TeV) proton beam to hit the (tungsten) target and produce a parent (of a few TeV or so) short-lived particle (together with plenty of background in wide energy range). Shortly downstream of the target the parent short-lived particle decays and produces a tau lepton (or charmed particle) among other particles. The focusing crystal traps the lepton or charmed baryon (because it emerges from the *F* point) but does not trap the particles emerging from the target.

Notice that, if the crystal would have a plain entry face (no focusing geometry) then it traps (in addition to a few short-lived particles) a vast amount of all kinds of particles mass-produced in the target by primary protons. The focusing entry face of a crystal is the key idea in this method of making beams of short-lived particles (tau leptons and charmed particles) as clean as possible.

With the principle understood, let us understand the limitations. Crystal traps a particle if it comes within a Lindhard angle $\theta_L$ which is e.g. 5 μrad/$E^{1/2}$ (TeV) in Si(110) planes at energy $E$ [13]. Therefore, the focus is not a precise point, but has a transverse size of $\pm\theta_L L$, where the focal length $L$ is as set by the crystal design (e.g. it was 0.5 to 4 m in the crystal focusing experiments at IHEP [13]). In Fig. 1, $L$ is the distance between points $F$ and $B$.

The typical angle $\theta_D$ under which the decay products emerge from $F$ is set by the relativistic factor $\gamma$ of the parent particle: $\theta_D = 1/\gamma$. Therefore, the longitudinal size of the focal spot equals $\theta_L L/\theta_D = \gamma\theta_L L$. Only the particles emerging from this focal spot of the length $\gamma\theta_L L$ and of the width $\pm\theta_L L$ can be trapped by the crystal (Fig. 2).

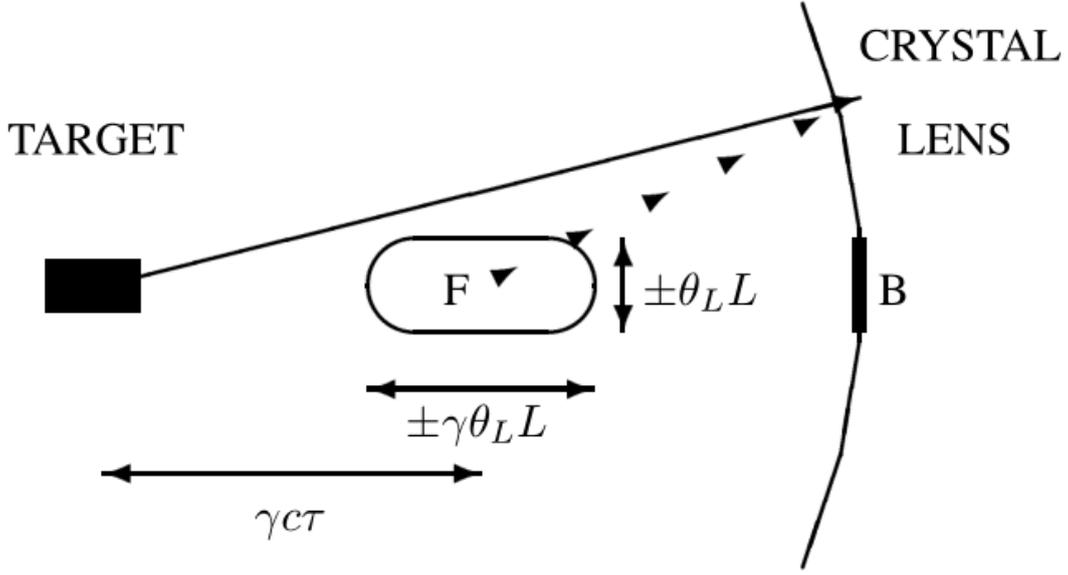

**Fig. 2** Schematic of a target and the focal spot dimensions. The "blind spot" is shown (B).

A particle with relativistic factor $\gamma$ and the lifetime $\tau$ has a mean decay path of $\gamma c\tau$. The discussed scheme makes sense only if $\gamma c\tau$ is on the order of or larger than $\gamma\theta_L L$. Hence, our technique can handle particles with $c\tau$ as small as $\theta_L L$. For the short-lived particles (with beauty or charm, or $\tau$ leptons etc.), $c\tau$ varies from 30 to 500 μm. The minimal $\pm\theta_L L$ already obtained in the IHEP Protvino experiment at 70 GeV was $\pm 20$ μm [13].

At higher energy $E$ the technique resolution $\theta_L L$ improves like $E^{-1/2}$. Moreover, $L$ is defined by just a practical convenience. Again, $\theta_L$ can be chosen smaller or bigger from different crystal planes or axes. E.g., using $L=0.2$ m at 1000 GeV gives the minimal resolution of ~1 μm. If one doesn't care on fine resolution for $c\tau$ but wish to increase the focal spot just to increase the working space, this is easy to do by means of a greater $L$.

The angular acceptance of the crystal should be about $\pm 1/\gamma$. The "blind spot" angular range in the crystal center is set by the focal spot transverse size $\theta_L L$ and the target-to-focus distance, about $\gamma c\tau$. The "blind" range is then $\pm\theta_L L/\gamma c\tau$. Its ratio to the crystal angular acceptance, i.e. the crystal inefficiency, equals $\theta_L L/c\tau$. As from above, this inefficiency is on the order of 1%.

We can also estimate the rejection factor for the particles incident beyond the Lindhard angle. They may be trapped in crystal through the scattering processes only, so called feed-in or volume capture. The probability of volume capture as measured at 70 GeV is on the order of 0.1% [13]. Therefore only ~0.1% of the background is not rejected by the crystal and hence mixed up with the "particles of interest" (our tau leptons). This probability decreases with energy like $R/E^{3/2}$ [13] where $R$ is the crystal bending radius. We estimate it to be on the order of 0.02% at 1 TeV. That is, the rejection factor for unwanted particles is about 5000.

Unfortunately, a high-efficiency trapping and bending has been demonstrated so far for positive particles mostly. Therefore, a major drawback of the discussed scheme (as well as of any other application of bent crystals) is its restriction to mostly positive particles produced in decay.

In principle, in this way one can trap any (positively charged) decay products of a parent short-lived particle. Charmed particles and τ leptons are among these decay products. Here, as one of applications for the presented idea, we suggest to make a beam of τ leptons and charmed mesons and baryons, similarly to kaon beams exploited at particle accelerators over many decades.

The considered method can be selective in $c\tau$. Suppose, your typical parent short-lived particle has γ =1000. If the focal point $F$ is about 500 mm from the tungsten target, then you trap the decay products of a parent particle with $c\tau$ of about 500 μm. If you place it at 100 mm from the target, then you sample all the cases from 100 μm and up to 500 μm. If you move the focal point $F$ farther/closer from/to the tungsten target, then you can handle particles with greater/smaller $c\tau$.

We conclude that the proposed technique can trap the decay products of the short-lived particles and bend them toward an experimental set-up. The rejection factor for the particles produced in the target is ~5000. The technique can handle particles with $c\tau$ down to ~1 μm and be used to produce beams of short-lived particles.

While the beams of strange particles (kaons) have served the physicists many decades, now we could arrange beams of charm and beauty particles and tau leptons. They can be made even in the ring over centimetre distance without supply of electrical power or cooling, with no expense for building the external beam line, and possibly using the collider detectors.

Notice that modern kaon beams are only a few percent kaons, the rest being pions, protons etc. How rich in tau leptons (charmed particles) can be the tau lepton (or charm) beam formed by a channeling crystal should be a subject of detailed optimization based on realistic simulations.

Accelerator scientists can collect the short-lived particles, focus them, and extract them from the mess of production region. They can suppress the background particles by many orders of magnitude in order to enrich the beam with short-lived particles. They can form the beam of short-lived particles and bend it onto experimental setup - and all of it over a matter of centimetres. Accelerator scientists can learn these opportunities and particle physicists can learn how to benefit from it.